
\font\tenrm=cmr10
\baselineskip=10pt
\magnification=1200
\tenrm
\baselineskip=2\baselineskip
\parindent=20pt
\hsize=15.5truecm
\vsize=24truecm
\bf
\centerline {QUANTUM GROUPS AND JORDAN STRUCTURES \footnote{$\dagger$}{ to
appear in {\it Bull. Univ. Politehnica Appl. Math. Sect. - Bucharest}}}
\vskip\baselineskip
\bigskip
\centerline {R. Iordanescu$^*$ and P. Truini$^{**}$}
\it
\centerline{$^*$ Institute of Mathematics of the Romanian Academy}
\centerline{P.O. Box 1-764, 70700 Bucharest - Romania}
\centerline{$^{**}$ Dipartimento di Fisica e Istituto Nazionale di Fisica
Nucleare}
\centerline{v. Dodecaneso 33, 16146 Genova - Italia}
\tenrm
\vskip\baselineskip
\bigskip\noindent
\hrule
\medskip\noindent
{\bf Abstract.} This paper is meant to be an informal introduction to
Quantum Groups, starting from its origins and motivations until the
recent developments. We call in particular the attention on the newly
descovered relationship among quantum groups, integrable models and
Jordan structures.
\hfil\break
\hrule
\vskip1.5truecm\noindent
\hbox{1. INTRODUCTION \hfill}
\bigskip
Quantum groups represent at present one of the most important topics of
mathematics and of mathematical physics.

The principal aim of the paper is to make the non-expert reader acquainted
with this fast developing field and motivate the investigation towards the
relationship with another important, though much earlier, branch of both
mathematics and mathematical
physics: that of Jordan structures. In the spirit of the word {\it quantum},
to denote a construction which includes the classical case in the appropriate
limit, we would like to point out that a {\it quantum} analog of
the profound link existing at the classical level between Lie and Jordan
structures is worth exploring.

We take this opportunity to firstly give a short introduction to quantum
groups. This is done in order to make our contribution understandable,
and also to introduce in the Romanian
mathematical literature this important and exciting topic in which a lot of
mathematicians and mathematical physicists around the world are now working.

The reader can complete the information given below by using, for instance,
the surveys by Biedenharn [1,2], Dobrev [3], Drinfeld [4], Faddeev [5],
Kundu [6], Majid [7], Ruiz-Altaba [8], Smirnov [9], Takhtajan [10]
- used also by us here - and, for an exhaustive information, the papers
referred
therein.
\bigskip
\hbox{2. HISTORICAL ROOTS AND THE BASIC APPROACH\hfill}
\bigskip
In this section we would like to outline the main ideas
which motivated many researches, from different fields, to work on topics
related to quantum groups and try to explain why quantum groups relate
many branches of theoretical physics and mathematics.

The origin of quantum groups lies in the search for the mathematical background
of the quantum inverse scattering method (QISM), a technique for obtaining
exact solutions for integrable quantum field theories in 1+1 dimensions and
classical models of statistical mechanics in dimension 2.
Such solutions belong to a class of models commonly referred to as
{\it exactly solved models}, which have a long history in theoretical
and mathematical physics, starting fom the work of Bethe, whose technique
for diagonalizing the Hamiltonian in one dimensional quantum spin systems
is known as the Bethe Ansatz Method, through the work of Onsager on two
dimensional lattice models of classical statistical mechanics (method of
commuting transfer matrices), to the more recent work of C.N. - C.P.
Yang, Baxter, and A.B.- Al.B. Zamolodchikov in field theory, and many
others.

Since all these methods share the feature of working in low dimension, let
us spend few words to recall why low-dimensional Physics has had
a great impulse in the last few years. Certainly an important role in this
respect is played by string theory: the 2 dimensional world sheet, namely the
analogue for the string of the the world-line for a relativistic
point particle, is strictly related to the
4-dim world (usually called the target space). It happens that the
ultraviolet limit of certain exactly solved models are Conformal Field
Theories which in turn describe string theories.
Another strong physical motivation comes from Quantum Chromo-Dynamics (QCD):
it has been argued that the
scattering process at high energies occurs essentially in dimension 2.
Moreover, to remain in the framework of quantum field theories,
some of the characteristic features of Yang-Mills theories in dimension 4,
like asymptotic freedom and dimensional transmutation, are also shared by the
non-linear sigma model in dimension 2.

These are not the only reasons for working in low dimension.
There is a branch of Physics in which low dimensional phenomena are observed
and studied: Condensed Matter Physics. Superconductivity at high-temperature
and quantum Hall effect are the most significant phenomena in this respect,
but we may also cite technological applications like silicon chips and plastic
films.

Beyond all this, low dimensional Physics has provided us with solvable models,
among them the first complete examples of field theory with non-trivial
interaction.
This is by itself a good reason for such an investigation: models are so
inherent to Physics!

With this variety of interests in low dimensional Physics, there is no
wander why a theory like QISM, which proved so powerful in yielding exact
solutions for a broad class of 2-dimensional models, has attracted so much
attention.

To understand how quantum groups are relaterd to QISM and what is the origin
itself of the term "quantum group", due to Drinfeld [4], one should
start from the fact that the QISM recovers
the classical inverse scattering method
(ISM) in the limit $h\to 0$, $h$ being the {\it quantum} parameter of the
theory, while in the same limit the quantum group underlying QISM goes to
the Poisson-Lie group underlying ISM. Let us briefly explain what is meant
by this.

The ISM is about 25 years old and has been successful in solving certain
types of classical field theories and Hamiltonian mechanics, like the ones
formulated in terms of the sine-Gordon equation, the non-linear
Schr\"odinger equation, the Korteweg de Vries equation, famous for its soliton
solution and now fundamental in the matrix models of non-perturbative
string theory (KdV hierarchy), the Toda systems, etc.
All these models are based on highly non-linear equations and
the key of the ISM is to introduce an auxiliary space and new operators
(Lax pairs) in terms of which one can write a linear equation equivalent to
the original one. By doing so one finds the solution to a typical inverse
problem analog to that of reconstructing a potential from the scattering
data. This is achieved using the technique of the commuting transfer matrices,
dating back to Onsager (since 1944).
We do not enter into details, since this is
beyond the scope of our paper, but what we have just mentioned is a crucial
point in the history of quantum groups. The method of commuting tranfer
matrices
leads to solvability by building a number of conserved quantities {\it in
involution}
equal to the number of degrees of freedom, thus ensuring integrability by
Liouville's theorem. Here "in involution" means that all these integrals
of motion, taken pairwise, have vanishing Poisson bracket; in the quantum
case this translates to being mutually commuting.
The main equation in the Hamiltonian formulation of ISM
is (in the discrete case):
$$\{ L_n(\lambda ) , L_m (\mu ) \} =
  [r(\lambda - \mu ), L_n(\lambda ) \otimes L_n (\mu ) ] \delta _{nm} $$
where $L$ is a Lax operator, that is a matrix -say $N\times N$-
depending on the classical fields, the curly bracket is the Poisson bracket,
$\{ L_n(\lambda ) , L_m (\mu ) \}$ is a $N^2\times
N^2$-matrix with elements $\{ L_n(\lambda )_{ij}, L_m (\mu )_{kl} \}$, and
$r$ (commonly called {\it classical r-matrix}) satisfies the classical
Yang-Baxter equation (YBE), leading quite directly to integrability:
$$[r_{12}(\lambda - \mu ), r_{13}(\lambda - \nu )]
+ [r_{12}(\lambda - \mu ), r_{23}(\mu - \nu )]
+ [r_{13}(\lambda - \nu ), r_{23}(\mu - \nu )] = 0 \eqno(2.1)$$
where, if $r$ acts as a matrix on $V\otimes V$, $r_{\alpha \beta}$ acts
on $V\otimes V\otimes V$ as $r$ on the $\alpha$-th and $\beta$-th components
and as the identity on the remaining one.

If $r(\lambda )\ \in \ \hbox{End}(V\otimes V)$ and $V$ is an $N$-dimensional
complex vector space, then we can regard the classical YBE as being
formulated in terms of the Lie algebra structure of End($V$). This suggests
to recast ISM in a more general algebraic structure. Let $\cal G$ be a Lie
algebra and $r(\lambda )$ be a ${\cal G} \otimes {\cal G}$ - valued function,
namely $$r(\lambda ) = r^{ij}(\lambda ) X_i \otimes X_j $$
wher $r^{ij}(\lambda )$ is a complex valued function.
Then
$$[r_{12}(\lambda ),r_{23}(\mu )] = r^{ij}(\lambda )\ r^{k\ell }(\mu )
X_i \otimes [X_j,X_k] \otimes X_\ell$$
so that each term of the classical YBE belongs to ${\cal G}\otimes
{\cal G}\otimes {\cal G}$.
The solution of the classical YBE, written in this way, is a {\it universal}
solution, in the sense that for each triplet of representations
$\{ \rho _\alpha , V_{\alpha} \}$ ($\alpha = 1,2,3$) the representation
$(\rho _\alpha \otimes \rho _\beta )(r_{\alpha \beta}(\lambda ))$ yields
a matrix solution of the classical YBE. For complex simple Lie algebras,
the solutions of the classical YBE have been studied by Belavin and  Drinfeld.
In particular Drinfeld has shown that if $G$ is a simple and simply connected
Lie group, then $G$ is a Poisson-Lie group, namely a Lie group with a
compatible Poisson structure built on the functions defined on the group
manifold, if and only if its Lie algebra
is a Lie bialgebra (for the concept of bialgebra see below). We may say
that the Lie bialgebras are the mathematical background for the theory
of classical integrable models (which is essentially soliton theory
in such a low dimension).

{\it Remark 2.1.} We like to point out that very recently Boldin, Safin and
Sharipov [11] proved a surprising connection between Tzitzeica surfaces and
ISM. The transformation that generates the family of such surfaces
found by Tzitzeica [12] in 1910 and its slight generalizations obtained by
Jonas [13,14] in 1921 and 1953 are known in the modern
literature on integrable equations as Darboux or B\"acklund tranformations.
They are used to construct the soliton solutions starting
from some trivial solution of the equation $u_{xy} = e^u - e^{-2u}$.
It is worth mentioning
that the paper [12] by Tzitzeica seems to be the first in the world where
the equation $u_{xy} = e^u - e^{-2u}$ (the nearest relative of the sine-Gordon
equation $u_{xy} = {\rm sin} u$) was considered.

Drinfeld's motivation for introducing quantum groups was to {\it quantize}
the structure underlying classical integrable models and classical YBE in
the sense of finding a quantum analog which recovers the classical theory
in the limit for a quantum parameter going to zero. This program led
him to quasi-triangular Hopf algebras in which a universal $R$-matrix is
defined satisfying the quantum Yang-Baxter equation (QYBE) (see below for
the definition) in each
representation of the algebra. The goal of quantizing the whole structure
rather then particular solutions of the classical YBE has thus been achieved.

The quantum $R$-matrix which in the limit goes to the classical $r$-matrix
$r(\lambda )\ \in {\cal G} \otimes {\cal G}$ (solution of the classical YBE)
lives in $U_q({\cal G})\otimes U_q({\cal G})$, $U_q({\cal G})$ being a
deformation of the universal enveloping algebra of $\cal G$. After all
$U_q({\cal G})$ is the natural candidate for the purpose, but we would
like to emphasize that one should not look only at the algebra aspect
of the theory: the coalgebra structure plays a crucial role from both
mathematical and physical standpoints [2] and both algebra and coalgebra are
deformed in general.

Let $V$ be a complex vector space and let $R(\lambda)$, $\lambda \in {\bf C}$,
be an End$_{\bf C}(V\otimes V)$-valued function, then the equation
$$R_{12}(\lambda - \mu) R_{13} (\lambda ) R_{23} (\mu ) \ = \
R_{23} (\mu ) R_{13} (\lambda ) R_{12}(\lambda - \mu ) \eqno(2.2) $$
is called Quantum Yang-Baxter Equation.

\noindent
{\it Comment.} It is possible to consider more general dependence of the
$R$-matrix on the spectral parameters and solutions have been found for
different cases.

We now want to show that this
equation is the key for the integrability of certain 1+1 dimensional quantum
systems as well as two dimensional classical statistical systems (the
equivalence between the two being one of the most interesting connections
between different branches of low dimensional physics).

{\it Remark 2.2}.
In the theory of quantum groups, or quasitriangular Hopf algebras,
a central role is played by the universal $R$-matrix satisfying (in all
representations of the Hopf algebra) an equation like (2.2) with no dependence
on the parameters $\lambda , \ \mu$, usually called {\it spectral parameters}.
Also the equation with no parameter dependence is referred to in the literature
as QYBE and to distinguish the two it is common to add "with spectral
parameter" when referring to (2.2).
In the QISM the spectral parameter is needed
for building up the generating functional for commuting conserved quantities.

For sake of simplicity, let us focus our attention on a vertex model
of statistical mechanics. We consider a two dimensional square lattice
with $N\times M$ points connected by bonds which can take $n$ possible
values. The {\it partition function} $Z$ gives the probability attached to
each configuration of the whole structure. In order to built it one
starts by setting up the Boltzmann weights at each vertex.
If the vertex has bonds $ijk\ell$ then the weight is denoted by
$R^{i\ k}_{\ j \ \ell}$ and can be arranged in a matrix form. In physical
applications the $R$-matrix should satisfy certain conditions, like
unitarity, symmetry, positivity
(the same for 1+1 quantum field theories) and very often periodic boundary
conditions are assumed and the limit $N,\ M \to \infty$ is eventually taken.
Going back to the construction of the partition function,
once the local properties of the model are set up, one looks for the
probability for a certain configuration of $N$ lattice points in a row.
Assuming periodic boundary conditions this is given by the {\it transfer
matrix}
$\tau = tr (\prod _{i=1}^N R^{(i)})$
(the trace appearing because of periodicity).
Finally the partion function is obtained by repeating the procedure on the
$M$ strings:  $Z= tr(\prod ^M \tau)$.
The model is exactly solvable if $R$ depends
on a parameter $\lambda \in {\bf C}$ such that eq.(2.2) is satisfied. Infact,
if this is the case, one can introduce an infinite set of commuting operators
as follows. First define the monodromy matrix:
$$T(\lambda ) ^{i \ \ \alpha _1\alpha _2 \dots \alpha _N}
_{\ j \ \ \beta _1\beta _2 \dots \beta _N} =
R(\lambda )^{i\ \alpha _1}_{\ k_1 \ \beta _1}
R(\lambda )^{k_1\ \alpha _2}_{\ k_2 \ \beta _2} \dots
R(\lambda )^{k_{N-1}\ \alpha _N}_{\ j \ \beta _N} $$
that can be viewed as an $n\times n$-matrix with values in End$(V^N)$ whose
trace $tr(T(\lambda )) \ \in \hbox{End}(V^N)$ is the transfer matrix (the
partition function, we recall, is the trace in $V^N$ of the transfer matrix).
The starting point of the QISM is the relation
$$R(\lambda )^{i\ k}_{\ m \ n} T(\lambda ')^m_{\ j} T(\lambda ")^n_{\ \ell}
 \ = \  T(\lambda ")^k_{\ n} T(\lambda ')^i_{\ m}
R(\lambda )^{m\ n}_{\ j \ \ell}$$
which follows from the definition of $T$ and the QYBE. This relation implies
the commutativity of the transfer matrices
$$ \left[ tr(T(\lambda ')), tr(T(\lambda ")) \right] \ = \ 0\; ,\qquad
\hbox{for all }  \lambda ' , \lambda " \ \in {\bf C} $$
hence the existence of an infinite number of conserved quantities, typically
regarded as the expansion coefficients ln$T(\lambda ) = \sum_{n=1}^\infty
C_n \lambda ^n$.

For quantum field theories in $1+1$ dimension the argument is essentially
the same with the change that one of the dimensions is time. This leads to
consider on a different footing the two spaces in which the bonds take values
instead of having two copies of the same space $V$ (an $n$-dimensional space)
both "horizontally" and "vertically" along the lattice.
For instance one may think
of the vertical space as the space of operators describing the time evolution
of the quantum mechanical system (possibly an infinite-dimensional space); the
$N$ horizontal points represent the discretization of space and the
continuum limit $N\to \infty $ yields integrable quantum field theories.
If we denote by $\bf V$ the vertical space to distinguish it from $V$, the
horizontal space, then $R \in$ End$(V\otimes {\bf V})$ is a Lax operator $L$
and the QYBE reads
$$R_{12}(\lambda -\mu ) L_1(\lambda ) L_2(\mu )\ = \ L_2(\mu ) L_1(\lambda )
  R_{12}(\lambda -\mu )  $$
where $R_{12}$ is a standard (i.e. numerical) matrix.
If such a relation holds at each lattice point $(i)$ and the
"ultralocality" condition
$$[ L^{(i)}(\lambda ) , L^{(j)}(\mu ) ] \ = \ 0 \; , \qquad i\ne j$$
is satisfied then one can pursue the program as above: construct the monodromy
matrix by multiplying the Lax operators and from it the commuting transfer
matrices which lead to integrability.

We now introduce formally quasi-triangular Hopf algebras (quantum groups
[1,15,16,17]).
The property of quasi-triangularity, in particular, involves the definition
of the universal $R$-matrix satisfying the Quantum Yang-Baxter equation
without spectral parameters. The process of {\it dressing} such an $R$-matrix
with parameters is called {\it Yang-Baxterization}, [18,19],
and is not treated in the present paper.

\noindent
\it Note. \tenrm The mathematics of quantum groups was also studied by Lusztig
[20,21], Rosso [22,23], Verdier [24].

Hopf Algebras involve both an algebric and a coalgebric structure.
Let us start by showing the algebric structure in the following definition.

\noindent
\it Definition 2.1. \tenrm If $\cal G$ is a complex Lie algebra, then the
\it extended enveloping algebra \tenrm $U_q ({\cal G})$
of the universal enveloping
algebra $U({\cal G})$ is the associative algebra over $\bf C$ with generators
$X^\pm_i, H_i, i = 1, 2, ...r$ = rank (${\cal G}$) and with relations
$$[H_i, H_j] = 0\ \ \ ,\ \ \ [H_i, X_j^\pm] = \pm a_{ij} X_j^\pm\ \ ,
\eqno(2.3)$$
$$[X_i^+, X_j^-] \ =\ \delta_{ij} \ \ {{q_i^{H_i/2} - q_i^{-H_i/2}}
\over {q_i^{1/2} - q_i^{-1/2}}}\ =\ \delta_{ij} [H_i]_{q_i}\ \ ,\ \ q_i =
q^{(\alpha_i, \alpha_i)/2}\ \ ,\eqno(2.4)$$
$$\sum_{k=0}^n \ (-1)^k {n \choose k}_{q_i} (X^\pm_i)^k X_j^\pm (X^\pm_i)^{n-k}
= 0\ \ ,\ \ i \not= j\ ,\eqno(2.5)$$
where $(a_{ij}) = \left(2 (\alpha_i, \alpha_j)/(\alpha_i, \alpha_i)\right)$ is
the Cartan matrix of ${\cal G}$, $( \;,\;)$ is the scalar product of the roots
normalized or that for the short simple roots $\alpha$ we have
$(\alpha, \alpha) = 2 \ ,\ n = 1 - a_{ij}$,
$${n \choose k}_q := {{[n]_q !} \over {[k]_q ! [n - k]_q !}}\ \ ,\ \
[m]_q ! := [m]_q [m - 1]_q ... [1]_q\ \ , \eqno$$
$$[m]_q\ :=\ {{q^{m/2} - q ^{-m/2}} \over {q^{1/2} - q^{-1/2}}}\ =\ {{sh
(mh/2)}
\over {sh (h/2)}}\ =\ {{\sin (\pi m \tau)} \over {\sin (\pi\tau)}}\ \ ,
\ \ q = e^h = e^{2\pi i \tau}, h, \tau \in {\bf C}, \eqno$$
$$q_i^{a_{ij}} = q^{(\alpha_i, \alpha_j)} = q_j^{\alpha_{ij}}\ \ .\eqno$$

\noindent
{\it Comment.} For a deeper analysis of the concepts involved in Definition 2.1
and in particular for an exhaustive discussion on the ring of functions of the
Cartan generators necessary in the construction of the extended enveloping
algebra we refer to Truini and Varadarajan [25].
\medskip\noindent
\it Remark 2.3. \tenrm The above construction works also when ${\cal G}$
is an affine Kac-Moody algebra (see Drinfeld [4]).

\noindent
\it Convention\tenrm. The subscript $q$ in $[m]_q$ will be omited if no
confusion can arise.

\noindent
\it Comments\tenrm. The extended enveloping algebra $U_q ({\cal G})$
is the "$q$-deformation" of the algebra $U({\cal G})$.
Definition 2.1. may be used also for real forms (introducing the appropriate
$\ast$-involutions) namely, where $q \in {\bf R}$ for the real compact forms
(via the Weyl unitary trick) - e.g., for the classical compact algebras $su(n),
so(n), sp(n)$ - while for $|q| = 1$, it may be used if ${\cal G}$ is replaced
by
its maximally split form - e.g., for the classical complex Lie algebras
these forms being $sl(n, {\bf R}), so(n, n), so(n+1, n), sp(n, {\bf R})$.

\noindent
\it Remark 2.4. \tenrm For $q \to 1 (\hbar \to 0)$ one recovers the standard
commutation relations from (2.3) and (2.4), and Serre's relations from (2.5) in
terms of generators $H_i, X_i^\pm$ (for the sense in which the limit is to be
understood see Drinfeld [4]).

We recall that the Serre relations allow to
define the universal enveloping algebra of a Lie algebra using only the
simple roots and the Cartan matrix. It is a classical result that one
thus gets an equivalent structure to the one obtained by taking the quotient
of the tensor algebra with generators corresponding to all the roots by the
full set of commutation relations (see Varadarajan [26]). There are two cases
in which the explicit introduction of the generators associated to non-simple
roots is convenient. One is the "q-analogue" of the Poincar\' e-Birkhoff-Witt
theorem (which selects a representative element in each equivalence class
of the quotient). The second is the {\it universal R-matrix} whose
representations are related through Yang-Baxterization
to the "physical $R$-matrix" which motivated the whole subject. Let us
thus spend a few words on the generators associated to non-simple roots.

\noindent
\it Conventions and notations. \tenrm The Cartan subalgebra, spaned by $H_i$,
will be denoted by $\cal H$, while the subalgebras spaned by $X_i^\pm$ will be
denoted by ${\cal G}^\pm$. We have the standard decompositions
$${\cal G} = {\cal H} \displaystyle \oplus_{\beta \in \Gamma}
{\cal G}_\beta = {\cal G}^+
\oplus {\cal H} \oplus {\cal G}^-\ , \ {\cal G}^\pm\ =\
\displaystyle \oplus_{\beta \in \Gamma ^\pm}
\ \ {\cal G}_\beta\ \ , \eqno$$
where $\Gamma = \Gamma ^+ \cup \Gamma ^-$ is the root system of $\cal G$ and
$\Gamma ^+, \Gamma ^-$ the sets of positive, negative roots, respectively. Let
us recall that the $H_i$'s correspond to the simple roots $\alpha_i$ of
$\cal G$, and if
$\beta = \sum_i n_i \alpha_i$, then to $\beta$ correspond $H_\beta = \sum_i
n_i H_i$. The elements of $\cal G$ which span ${\cal G}_\beta$ will be
denoted by $X_\beta$.
These Cartan-Weyl generators are normalized, so that we have
$$[X_\beta, X_{-\beta}] = [H_\beta]_{q\beta}\ \ {\rm{for}}\ \  \beta \in
\Gamma ^+\ \ ,\ \ q_\beta = q^{(\beta, \beta)/2}\ \ .\eqno(2.6)$$

\noindent
\it Remark 2.5. \tenrm Instead of $H_i$, some authors prefer to use $K_i^\pm$
defined by
$$K_i^\pm : = q_i^{\pm H_i/2}\ \ , \eqno$$
and then (2.3) and (2.4) become
$$K_i, K_i^{-1} = K_i^{-1} K_i = 1,\; [K_i, K_j] = 0, \; K_i X_j^\pm K_i^{-1} =
q_i^{\pm a_{ij}/2} X_j^\pm\ \ ,\eqno (2.3')$$
$$[X_i^+, X_j^-] \ =\ \delta_{ij} \ \ {{K_i - K_i^{-1}}
\over {q_i^{1/2} - q_i^{-1/2}}}\ \ .\eqno(2.4')$$

\noindent
\it Remark 2.6. \tenrm One can use, following Rosso [27], instead of
$X_i^\pm$, the generators $E_i$ and $F_i$ defined as follows
$$E_i := X_i^+ q_i^{-H_i/4}\ =\ X_i^+ K_i^{-1/2}\ \ ,\ \ F_i :=
X_i^- q_i^{H_i/4} \ =\ X_i^- K_i^{1/2}\ \ .\eqno(2.7)$$

\noindent
\it Definition 2.2. \tenrm An associative algebra $\cal A$ with unit
$1_{\cal A}$ is called
a \it bialgebra \tenrm if there exist two homomorphisms $\Delta$ and
$\varepsilon$, called, \it comultiplication \tenrm and \it counit\tenrm,
respectively, such that
$$\Delta \ : \ {\cal A} \to {\cal A} \otimes {\cal A}\ \ ,\ \
\Delta (1_{\cal A}) = 1_{\cal A} \otimes 1_{\cal A}   $$
and
$$\varepsilon \ : \ {\cal A} \to {\bf C}\ \ ,\ \ \varepsilon (1_{\cal A}) =
1\ \ , $$
the comultiplication $\Delta$ satisfying the axiom of coassociativity
$$(\Delta \otimes Id) \circ \Delta\ \ =\ \ (Id \otimes \Delta) \circ \Delta\ \
,
$$
where both sides are maps ${\cal A} \to {\cal A} \otimes {\cal A}
\otimes {\cal A}$, the two homomorphisms
fulfilling
$$(Id \otimes \varepsilon) \circ \Delta = i_1\ \ \ ,\ \ \ (\varepsilon \otimes
Id) \circ \Delta = i_2 $$
as maps ${\cal A} \to {\bf C} \otimes {\cal A}, {\cal A} \to {\cal A} \otimes
{\bf C}$, respectively, where
$i_1, i_2$ are the maps identifying $\cal A$ with ${\cal A} \otimes {\bf C},
{\bf C} \otimes {\cal A}$, respectively.

\noindent
\it Definition 2.3. \tenrm A bialgebra $\cal A$ is called a \it Hopf algebra
\tenrm if there exists an algebra antihomomorphism $S$ - called \it
antipode \tenrm - such that
$$S \ : \ {\cal A} \to {\cal A}\ \ ,\ \ S (1_{\cal A}) = 1_{\cal A} $$
and
$$m \circ (Id \otimes S) \circ \Delta = i \circ \varepsilon\ \ ,$$
as maps ${\cal A} \to {\cal A}$, where $m$ is the usual product
in the algebra ${\cal A}$
(i.e. $m(Y \otimes Z) = YZ,\; Y, Z \in {\cal A})$,
and $i$ is the natural embedding of $\bf C$ into $\cal A$ (i.e.
$i(c) = c\ 1_{\cal A}\ , \ c \in {\bf C}$)

\noindent
\it Remark 2.7. \tenrm The antipode plays the role of an inverse although
there is no requirement that $S^2 = Id$.

\noindent
\it Remark 2.8. \tenrm Following Dobrev [3], we shall use also the \it opposite
comultiplication \tenrm $\Delta^\prime : = \sigma \circ \Delta$, where
$\sigma$ is the permutation in ${\cal A} \otimes {\cal A}$.
In case the antipode has
an inverse, then one uses also the \it opposite antipode \tenrm $S^\prime
: = S^{-1}$ (see also Drinfeld [4] and Jimbo [16]).

The comultiplication, counit, and antipode are defined on the generators
of $U_q ({\cal G})$ as follows:
$$\Delta (H_i) = H_i \otimes 1 + 1 \otimes H_i\ ,\ \Delta (X_i^\pm) = X_i^\pm
\otimes q_i^{H_i/4} + q_i^{-H_i/4} \otimes X_i^\pm\ \ ,$$
$$\varepsilon (H_i) = \varepsilon (X_i^\pm) = 0\ \ ,\eqno(2.8)$$
$$S (H_i) = - H_i\ ,\ S (X_i^\pm) = - q_i^{\hat\rho/2} X_i^\pm
q_i^{-\hat\rho/2} = - q_i^{\pm 1/2} X_i^\pm\ \ ,\eqno$$
where $\hat \rho \in {\cal H}$
corresponds to $\rho = {1\over 2} \sum_{\alpha \in
\Gamma ^+}\ \alpha\ ,\ \Gamma ^+$ being the set of positive roots, $\hat \rho
= {1\over 2} \sum_{\alpha \in \Gamma^+} H_\alpha$.

\noindent
\it Comment. \tenrm As it was remarked, for ${\cal G} = sl (2, {\bf C})$,
in the paper [28] by Sklyanin, and, in general, in the papers [4,15]
by Drinfeld and [16,17] by Jimbo, the algebra $U_q({\cal G})$
is a Hopf algebra.

\noindent
\it Remark 2.9. \tenrm The above formulae (2.8) hold also for $H_\beta,
X_{\pm\beta}$ from (2.6).

\noindent
\it Remark 2.10. \tenrm The opposite comultiplication and antipode from
Remark 2.8. define a Hopf algebra $U_q({\cal G})^\prime$ which is related to
$U_q({\cal G})$ by the relation $U_q({\cal G})^\prime = U_{q^{-1}} ({\cal G})$.

In terms of the generators $K_i^\pm, E_i, F_i$ from Remarks 2.5. and 2.6., the
relations (2.8) become
$$\Delta (K_i) = K_i \otimes K_i\ ,\; \Delta (E_i) = E_i
\otimes 1 + K_i^{- 1} \otimes E_i\ ,\; \Delta (F_i) = F_i \otimes K_i + 1
\otimes F_i\ \ ,$$
$$\varepsilon (K_i) = 1\ \ ,\ \ \varepsilon (E_i) = \varepsilon (F_i) = 0
\eqno(2.8')$$
$$S(K_i) = K_i^{- 1}\ ,\ S(E_i) = - K_i E_i\ ,\ S(F_i)
= - F_i K_i^{- 1}\ \ .$$
One can also rewrite Serre's relations (2.5) as follows
$$(ad_q E_i)^n (E_j) = 0 = (ad^\prime_q F_i)^n (F_j)\ \ ,\ \ i \not= j\ ,$$
where
$$ad_q : U_q ({\cal G}^+) \to End\ (U_q({\cal G}^+))\ ,\ ad_q = (L \otimes R)
(Id \otimes S) \Delta \ , \eqno$$
$$ad_q^\prime : U_q ({\cal G}^-) \to End\ (U_q({\cal G}^-))\ ,
\; ad_q^\prime = (L \otimes R)
(Id \otimes S^\prime) \Delta^\prime \ , \eqno(2.5')$$
and $L$ (resp. $R$) is the left (resp. right) multiplication.

\noindent
\it Remark 2.11. \tenrm $ad_q (E_i)$ acts as a twisted derivation.

\noindent
\it Definition 2.4. \tenrm A Hopf algebra ${\cal A}$ for which there exists an
invertible element $R$ in ${\cal A} \otimes {\cal A}$
- called \it universal R-matrix \tenrm
(cf. Drinfeld [4,15]) - which intertwines $\Delta$ and $\Delta^\prime$,
i.e.
$$R \Delta (Y) = \Delta^\prime (Y) R\ \ ,\ \ \hbox{for all}\ \ Y\ \ {\rm{in}}
\ \ {\cal A}\ \ ,
\eqno(2.9a)$$
and obeys also the relations
$$(\Delta \otimes Id) R\ =\ R_{13} R_{23}\ \ ,\ \ R = R_{.3}\ \ ,\eqno(2.9b)$$
$$(Id \otimes \Delta) R\ =\ R_{13} R_{12}\ \ ,\ \ R = R_{1.}\ \ ,\eqno(2.9c)$$
where the indices indicate the embeddings of $R$ into ${\cal A}
\otimes {\cal A} \otimes {\cal A}$
is called a \it quasi-triangular Hopf algebra\tenrm.

\noindent
\it Remark 2.12. \tenrm We have
$$(\varepsilon \otimes Id) R\ =\ (Id \otimes \varepsilon) R = 1_{\cal A}\
\ ,\ \ {\rm{and, moreover}}\ \ ,$$
$$(S \otimes Id) R\ =\ R^{-1} \ \ ,\ \ (Id \otimes S) R^{-1} = R\ \ . $$

\noindent
\it Definition 2.5. \tenrm A quasi-triangular Hopf algebra for which also
$\sigma R^{-1} = R$ holds is called a \it triangular Hopf algebra\tenrm.

\noindent
\it Definition 2.6. \tenrm From (2.9a) and one of (2.9b,c) it follows
$$R_{12}\ R_{13}\ R_{23}\ \ =\ \ R_{23}\ R_{13}\ R_{12}\ \ , \eqno$$
which is the Yang-Baxter equation for $R$ without spectral parameters.

\noindent
\it Comments\tenrm. The universal $R$-matrix was given explicitly for ${\cal G}
= sl(2, {\bf C})$ by Drinfeld [15], namely,
$$R = q^{H \otimes H/4}\ \sum_{n \geq 0}\ \ {{(1-q^{-1})^n q^{{{n(n-1)}\over
4}}} \over {[n]!}}\ \ \left(q^{{H \over 4}} X^+\right)^n \otimes
\left( q^{-{H\over 4}} X^-\right)^n\ \ .$$
This $R$-matrix is not in
$U_q (sl(2, {\bf C})) \otimes U_q (sl(2, {\bf C}))$, since it contains power
series involving $X^\pm$, but it is in some completion of it (in the h-adic
topology used by Drinfeld [4,15] $(q=e^h)$). This fact is valid for
the $R$-matrices of all $U_q ({\cal G})$.
Hopf algebras with such an $R$-matrix are
called by Drinfeld [4] \it pseudo quasi-triangular Hopf algebras\tenrm,
and by Majid [7] \it essentially quasi-triangular Hopf algebras\tenrm.
For ${\cal G} = sl (n, {\bf C})$, an explicit formula for $R$ was given by
Rosso
[27].
\bigskip
\hbox{3. SUBSEQUENT AND RECENT DEVELOPMENTS\hfill}
\bigskip
At the end of eighties other approaches to quantum groups were given. The
objects of these approaches - which can be called \it quantum matrix groups
\tenrm - are Hopf algebras in duality to quantum algebras.

\noindent
\it Definition 3.1. \tenrm Two Hopf algebras ${\cal A}$ and ${\cal A}'$
are said to
be in \it duality \tenrm if there exists a doubly nondegenerate bilinear form
$$<\ ,\ > : {\cal A} \times {\cal A}'
\to {\bf C}\ \ ,\ \ <\ ,\ > : (a, a') \to <a, a'>\ , $$
such that, for $a, b \in {\cal A}$ and $a', b' \in {\cal A}'$ the
following relations hold
$$<a, a'b' > = <\Delta_{\cal A} (a), a' \otimes b'>\ \ ,\ \
<ab, a'> = <a \otimes b, \Delta_{{\cal A}'} (a')>\ \ ,$$
$$<1_{\cal A}, a'> = \varepsilon_{{\cal A}'} (a')\ ,\
<a, 1_{{\cal A}'}>\ =\ \varepsilon_{\cal A} (a)\ ,
\ <S_{\cal A} (a)\ ,\ a'>\ =\ <a, S_{{\cal A}'} (a')>\ \ .$$
One of these approaches is due to Faddeev, Reshetikhin and Takhtajan [29,30]
and it is called "$R$-matrix approach". It is based on the main relation of the
quantum inverse scattering method. The quantum group matrices play the role
of the quantum monodromy matrices (with operator-value entries) of the
auxiliary linear problem and the Yang-Baxter equation is the compatibility
equation.

Another approach is that of Manin [31,32,33], who considers quantum groups
as symmetries of non-commutative, or quantum, spaces. The resulting
objects are the same as those of the first approach.

A third approach that we would like to mention here is that of Woronowicz
[34,35,36]. This approach also deals with same objects with some additional
structures since Woronowicz's starting point is the theory of $C^*$-algebras.

\noindent
\it Note\tenrm. Connections between the above mentioned approaches can be
found in the papers by Doebner, Hening and L\"ucke [37], Majid [7],
and Rosso [22].
\bigskip
At the end of eighties, Drinfeld [15], inspired by Knizhnik-Zamolodchikov
equations (see [38]), developed a theory of formal deformations and introduced
a new notion of "quasi-Hopf algebras".

The matrix quantum group approaches were recently developed, in particular
finding consistent multiparametric deformations. This is related to the
development by Wess, Zumino and collaborators of differential calculus on
quantum hyperplanes. The latter approach is actually an example of
non-commutative differential geometry which in opinion of Manin [39] is
different in spirit from that of Connes [40], although the exact relation
is not known yet. For a detailed review of the above mentioned developments we
refer the reader to \S\S 2-5 from Dobrev's paper [3].

Finally, we like to mention the very recent paper by Mack and Schomerus
[41] on "quantum field planes". These objects are generalizations of the
quantum planes which were studied by Manin, Wess and Zumino, and others,
and were generalized to the quasi-associative case by the same authors
[42]. Basically, the construction of quantum field
planes replaces the ground field $\bf C$ of quantum planes by the
non-commutative algebra of observables of the quantum field theory in the
local Lorentz frame.

\noindent
\it Comments\tenrm. An interesting open problem could be to reconsider
Mack-Schomerus construction in case when the algebra of observables is a
(non-commutative) Jordan algebra.

\noindent
\it Remark 3.1. \tenrm As a final note to this survey on the history of quantum
groups we want to emphasize that quantum groups are involved and studied in
many
different fields of mathematics and physics some of which we haven't even
mentioned so far. Among them are:
topological quantum field theories, 2-dimensional gravity and
3-dimensional Chern-Simons theory [43,44,45,7],
rational conformal field theory [46,47,48],
braid and knot theory [49,50], non-standard quantum statistics [51], quantum
Hall effect [52,53].

\bigskip
\hbox{4. RELATIONS WITH JORDAN STRUCTURES\hfill}
\bigskip
We would like to start by recalling here the new topic proposed by Truini and
Varadarajan at the end of their recent paper [54], namely: \it quantization
of Jordan structures\tenrm.

Let us explain why it is interesting to investigate this new topic.

It is known that the quantization of the Poincar\'e group is receiving
more and more attention by the physicists. The main reason of this attention
consists in its relationship with the non-commutative geometry of quantum
Minkowski space (see [55]). It is believed that the models of non-commutative
space-time, and their quantum symmetry groups, may provide a basis for building
a divergence free theory of elementary particles and their fields, including
gravitation, thus overcoming the difficulties arising out of the structure
of conventional space-time at small distances. Consequently, Truini and
Varadarajan (see [54, p.732]) intend to study quantizations of semidirect
products as Hopf algebras which maintain the classical picture of a space
and a set of transformations that act covariantly on it. But, this point
of view is related to the general problem of deforming Jordan structures.
Indeed, the algebra of Poincar\'e group can be represented as the semidirect
product of ${\cal L} \simeq sl (2, {\bf C})$
and the Jordan algebra $J$ (representing
the translations) of $2 \times 2$-Hermitian matrices over $\bf C$,
$${{\cal L}\ \ \ J \choose 0\ \ \ -{\cal L}^\dagger}\eqno$$
the action of the algebra of the Lorentz group on the translations being
determined by the commutator of the related $4 \times 4$-matrices.

Another motivation for the study of deformations of Jordan structures comes
from the conformal group, in which the Poincar\'e group is naturally imbedded
and which is a simple group, thus falling into their general theory of
universal
deformations. Similarly to the above representation of the Poincar\'e group,
the algebra of the conformal group can be written as
$$\pmatrix{\tilde {\cal L} &  J_1 \cr J_2 & -\tilde {\cal L}^+\cr} \eqno$$
where $J_1$ and $J_2$ are Jordan algebras of $2 \times 2$-Hermitian matrices
over $\bf C$ and $\tilde {\cal L} \simeq sl(2, {\bf C}) \oplus {\bf R}$. It is
easy to show that $(J_1, J_2)$ is a Jordan pair, $\tilde {\cal L}$
is the algebra of its automorphism group and the structural algebra of $J$.

It is known that a Jordan pair $V = (V^+, V^-)$ is a pair of modules acting
each other through a map $U_{x^+} x^-$ quadratic in $x^+$ and linear in
$x^-$ and obeying certain axioms which extend those of the quadratic
formulation of the theory of Jordan algebras (where $U_x y$ generalizes
$xyx$). Roughly speaking, a Jordan pair is "a pair of spaces acting on each
other like a Jordan triple system" - see McCrimmon [56, p.621].
On the other hand, let us point out that Jordan pairs are much more common
than what one would expect. So, there exists a 1-1 correspondence between
3-graded Lie algebras
$$L = L_1 \oplus L_0 \oplus L_{-1} \ \ \ ([L_i, L_j] \subset L_{i+j})\eqno$$
and Jordan pairs $V$ (plus the choice of $L_0$ between Der ($V$) and Inder
($V$)) - see, for instance, McCrimmon [56, p.622]. As it is known, many Lie
algebras are three-graded (e.g., $A(2), C(3), E(7)$).

In this
respect, it is interesting to point out that Okubo [57] has related triple
products, which are linearization of the quadratic maps, to the quantum
$R$-matrix and used the relationship to find solutions to the quantum
Yang-Baxter equation.

More recently, in a series of papers, Okubo [58,59,60,61] reformulated first
the Yang-Baxter equation as a triple product relation and then solved it
for triple systems called "orthogonal" and "symplectic" (see Okubo [58,59]).
A superspace extension of this work was given by Okubo himself in [60,61].

In a very recent paper, Okubo [62] gave some solution of Yang-Baxter equation
in terms of Jordan triple systems and of so-called "anti-Jordan triple
systems".
(see Definition 4.1. below).

\noindent
\it Definition 4.1. \tenrm Let $V$ be a $N$-dimensional vector space over a
field $\bf F$ and let $xyz : V \otimes V \otimes V \to V$ be a triple product
in $V$ satisfying the following conditions
$$z y x\ \ =\ \ \delta x y z\ \ ,\eqno (i)$$
$$u v (x y z)\ \ =\ \ (u v x) y z - \delta x (v u y) z + x y (u v z)\ \ ,
\eqno (ii)$$
where $\delta = \pm 1$. The case $\delta = 1$ defines the well known (linear)
\it Jordan triple systems \tenrm (see, for instance, Meyberg [63]),
while the case
$\delta = -1$ defines the \it anti-Jordan triple systems\tenrm.

\noindent
\it Note\tenrm. As it was noticed by Koecher [64], a glimps of Jordan triple
system was given by Gibbs (1839-1903) as early as 1881 (Collected Works,
vol.II, p.18) in a different setting.

\noindent
\it Comments\tenrm. Compare the definition of "anti-Jordan triple systems"
given by Okubo [62] (see Definition 4.1.) with the definition of
"anti-Jordan pairs" given by Faulkner and Ferrar in [65].

Okubo [62] considered $V$ endowed also with a bilinear non-degenerate form
$<x|y>$ satisfying
$$<y|x>\ = \epsilon <x|y> \qquad\epsilon = \pm 1 $$
\medskip
Let $R(\theta ) \in End(V)\otimes End(V)$ be the scattering matrix
with matrix elements $R^{dc}_{ab}$, defined by $R(\theta )e_a\otimes e_b =
R^{dc}_{ab} e_c \otimes e_d$, with respect to a basis $\{ e_j\} $ of $V$
and suppose that $R$ satisfies the QYBE
$${R}_{12} (\theta) {R}_{13} (\theta^\prime)
{R}_{23} (\theta^{\prime \prime}) =
{R}_{23} (\theta^{\prime \prime}) {R}_{13} (\theta^\prime)
R_{12}(\theta)  \eqno(4.1a)$$
with
$$\theta^\prime = \theta + \theta^{\prime \prime} \eqno(4.1b)$$
\medskip
Two $\theta$-dependent triple linear products $[x,y,z]_\theta$ and
 $[x,y,z]^*_\theta$ are defined in terms of the scattering matrix
 elements $R^{dc}_{ab} (\theta)$, by
$$\eqalign{\big[ e^c,e_a,e_b \big]_\theta &:= e_d R^{dc}_{ab}
(\theta)  \cr
\big[ e^d,e_b,e_a \big]^*_\theta &:= R^{dc}_{ab}
(\theta)e_c  \cr}$$
or alternatively by
$$R^{dc}_{ab} (\theta) =\ <e^d | \big[ e^c,e_a,e_b\big]_\theta>\ =\
<e^c | \big[e^d,e_b,e_a \big]^*_\theta> $$
where $e^d$ is given by
$$<e^d |e_c>\ = \delta^d_c \quad . $$
The QYBE (4.1a) can be then rewritten as a triple product
equation
$$\eqalign{\sum^N_{j=1} \big[ &v,[u,e_j,z]_{\theta^\prime} ,
[e^j,x,y]_\theta \big]^*_{\theta^{\prime \prime}} \cr
&= \sum^N_{j=1} \big[ u,[v,e_j,x]^*_{\theta^\prime} ,
[e^j,z,y]^*_{\theta^{\prime \prime}} \big]_\theta \quad . \cr}\eqno(4.2)$$
\smallskip

\noindent {\it Proposition 4.1}
Let $V$ be a Jordan or anti-Jordan triple system with
$\epsilon =1$ satisfying the following conditions
\item{i)} $<u|xvy>\ =\ <v|yux> $
\medskip
\item{ii)} $<u|xvy>\ =\ \delta <x|uyv>\ =\ \delta <y|vxu>$
\medskip
\item{iii)} $(y e^j x)v e_j = a\{ <x|v>y + \delta <y|v>x\}
+ b\ y\ v\ x $

\medskip

\item{(iv)} $(ye^jx) v(zue_j) -
(ye^jz)u(xve_j) = \alpha \{<v|x>zuy\  -
\ <u|z>xvy\}$

\item{   } $\qquad \qquad + \beta \{ <v|y>xuz \ -\ <u|y> zvx\  +\
<y|uzv>x$

\item{   } $\qquad \qquad - <y|vxu>z\} + \gamma \{
(yux)vz - (yvz) ux\}$

\noindent for some constants $a,\ b,\ \alpha,\ \beta, {\rm and}\
\gamma$.
Then,
$$[x,y,z]_\theta = P(\theta) \ y\ x\ z + B(\theta)
<x|y>z + C(\theta)<z|x>y $$
for $P(\theta) \not= 0$ is a solution of the QYBE (4.2) with
$${B (\theta) \over P (\theta)} = \delta \gamma + k \theta
\quad , \quad {C(\theta) \over P(\theta)} =
{\beta \delta \over k \theta} $$
for an arbitrary constant $k$, provided that we have either

\item{(i)} $\alpha = \beta = 0 \quad ,$  \hfill

\noindent or

\item{(ii)} $\alpha = \beta \not= 0 \quad , \quad b = -2 \gamma \quad
, \quad a = 2 \beta$ \quad . \hfill

\noindent {\it Remark 4.1}
The solution satisfies the unitarity condition
$$R(\theta)\ R(-\theta) = f(\theta) \ Id $$
where
$$f(\theta) = P(\theta) P(-\theta) \bigg\{ (a+\gamma^2) - (k \theta)^2
- {\beta^2 \over (k \theta)^2} \bigg\} $$
\medskip
\noindent {\it Proposition 4.2}
Let $V$ be the Jordan triple system defined on
the vector space of the Lie-algebra $u(n)$ by means of the product
$$x\ y\ z = x \cdot y \cdot z + \delta \ z \cdot y \cdot x
$$
the dot denoting the usual associative product in $V$ and let $<\; | \; >$
be the trace form. Then,
$$[x,y,z]_\theta = P(\theta) x\ z\ y + A(\theta) <y|z>x\ +
C(\theta) <z|x>y $$
for $P(\theta) \not= 0$ offers solutions of the QYBE (4.2) for
the following two cases:

\item{(i)} ${A(\theta) \over P(\theta)} = - {\lambda^2
e^{k \theta} - d \over \lambda (e^{k \theta}-d)}
\quad , \quad {C(\theta) \over P(\theta)} = -
{e^{k \theta} - \lambda^2 \over
\lambda (e^{k \theta}-1)}$ \hfill (4.3a)

\noindent where $d$ is either $\lambda^2$ or $-\lambda^4$ and $k$ is an
arbitrary constant, or

\item{(ii)} ${A(\theta) \over P(\theta)} = - \lambda \quad , \quad
{C(\theta) \over P(\theta)} = - {1 \over \lambda} \quad .$
\hfill (4.3b)

\noindent In both cases $\lambda$ is given by
$$\lambda = {1 \over 2}\ \left( n \pm \sqrt{n^2 -4}
\right) $$

\smallskip

\noindent {\it Remark 4.2}
The first solution Eq. (4.3a) satisfies both unitarity and crossing
 symmetry relations:

$$\eqalignno{&R (\theta) R (-\theta) =
C(\theta) C(-\theta) Id &(4.4a)\cr
\noalign{\vskip 4pt}%
&{1 \over P (\overline \theta)} \ [y,x,z]_{\overline \theta}
= {1 \over P(\theta)} \ [x,y,z]_\theta &(4.4b)\cr}$$
where $\overline \theta$ in Eq. (4.4b) is related to $\theta$ by
$$\theta + \overline \theta = {1 \over k}\
\log d \quad . $$
In view of these, the solution is likely related [66] to some exactly
solvable two-dimensional quantum field theory.

Finally we want to mention a result by Svinolupov [67] which is interesting
in the context of this paper. He considers systems of nonlinear equations
which, in a particular case, may be reduced to the nonlinear Schr\"odinger
equation and are therefore called generalized Schr\"odinger equations. A one
to one correspondence between such integrable systems and Jordan Pairs is
established. It turns out that {\it irreducible} systems correspond to
{\it simple} Jordan Pairs.

In our opinion the general setting in which one should consider the problem
of quantizing the Jordan structures is that of Jordan Pairs. We are currently
investigating this possibility with the belief that a quantum analog
of the classical link between Jordan and Lie structures would give
a deeper insight and reveal new aspects in the theory of quantum groups.
\vskip1.5truecm
\line{REFERENCES \hfil}
\bigskip
\item{1.} Biedenharn, L.C.: invited paper presented at the XVIII Int. Coll.
on Group Theoretical Methods in Physics, Moscow, USSR, June 4-9 1990
\item{2.} Biedenharn, L.C.: Int. J. Theor. Phys. {\bf 32} (1993) No.10, 1789
\item{3.} Dobrev, V.K.: Invited plenary lecture at the 22nd Iranian Mathematics
Conference, March 13-16 1991, Moshhad Iran
\item{4.} Drinfeld, V.G.: in Proc. Int. Congress of Math. Berkeley, California
, Academic Press, New York, {\bf 1}, 1986 p.798
\item{5.} Faddeev, L.D.: preprint ITP-SB-94-11 and HEP-TH 9404013 (1994)
\item{6.} Kundu, A.: in "Application of solitons in Science and Engeneering",
World Scientific, Singapore, 1994 (to appear)
\item{7.} Majid, S.: Int. J. Mod. Phys. {\bf A5} (1990), 1
\item{8.} Ruiz-Altaba, M.: preprint UGVA-DPT-1993-10-838 and HEP-TH 9311069
(1993)
\item{9.} Smirnov, F.A.: in "Introduction to Quantum Group and Integrable
Massive Models of Quantum Field Theory" World Scientific, Singapore, 1990 p.1
\item{10.} Takhtajan, L.A.: in "Introduction to Quantum Group and Integrable
Massive Models of Quantum Field Theory" World Scientific, Singapore, 1990 p.69
\item{11.} Boldin, A.Yu., Safin, S.S., Sharipov, R.A.: J. Math. Phys. {\bf 34}
(1993), No.12, 5801
\item{12.} Tzitzeica, G.: C.R. Acad. Sci. paris {\bf 150} (1910), 955
\item{13.} Jonas, H.: Ann. Mat. XXX (1921), 223
\item{14.} Jonas, H.: Math. Nachr. {\bf 10} (1953), 331
\item{15.} Drinfeld, V.G.: Alg. Anal. {\bf 1} (1989)
\item{16.} Jimbo, M.: Lett. Math. Phys. {\bf 10} (1985), 63
\item{17.} Jimbo, M.: Lett. Math. Phys. {\bf 11} (1986), 247
\item{18.} Jones, V.F.R.: Int. J. Mod. Phys. {\bf B4} (1990), 701
\item{19.} Basu Mallick, B., Kundu, A.: J. Phys. {\bf A25} (1992), 4147
\item{20.} Lusztig, G.: Advances Math. {\bf 70} (1988), 237
\item{21.} Lusztig, G.: {\it Introduction to Quantum Groups}, Birkh\"auser,
Basel, 1993
\item{22.} Rosso, M.: C.R. Acad. Sci. Paris {\bf 305} (1987), Serie I, 587
\item{23.} Rosso, M.: Commun. Math. Phys. {\bf 117} (1987), 581
\item{24.} Verdier, J.-L.: in S\'eminaire Bourbaki, No. 685 Ast\'erisque
{\bf 152-153} (1987), 305
\item{25.} Truini, P., Varadarajan, V.S.: Rev. Math. Phys. {\bf 5} (1993),
No.2, 363
\item{26.} Varadarajan, V.S.: {\it Lie Groups, Lie Algebras and their
Representations}, Prentice Hall Inc., Englewood Cliffs, N.J. 1974
\item{27.} Rosso, M.: Commun. Math. Phys. {\bf 124} (1989), 307
\item{28.} Sklyanin, E.K.: Uspekhi Mat. Nauk. {\bf 40} (1985), 214
\item{29.} Faddeev, L.D., Reshetikhin, N.Yu., Takhtajan, L.A.:
Alg. Anal. {\bf 1} (1989), 178
\item{30.} Takhtajan, L.A.: Adv. Stud. Pure Mat. {\bf 19} (1989), 435
\item{31.} Manin, Yu. I.: Ann. Inst. Fourier {\bf 37} (1987), 191
\item{32.} Manin, Yu. I.: Montreal Univ. preprint, CRM-1561 (1988)
\item{33.} Manin, Yu. I.: Commun. Math. Phys. {\bf 123} (1989), 163
\item{34.} Woronowicz, S.L.: Commun. Math. Phys. {\bf 111} (1987), 613;
Lett. Math. Phys. {\bf 21} (1991), 35
\item{35.} Woronowicz, S.L.: Publ. RIMS {\bf 23} (1987), 117
\item{36.} Woronowicz, S.L.: Commun. Math. Phys. {\bf 122} (1989), 125
\item{37.} Doebner, H.D., Hennig, J.D., L\"ucke, W.: in Proc. Quantum
Groups Workshop, Clausthal 1989, Eds. H.D. Doebner and G.D. Hennig, Lecture
Notes in Physics {\bf 370} (Springer-Verlag Berlin 1990) p.29
\item{38.} Knizhnik, V.G., Zamolodchikov, A.B.: Nucl. Phys. {\bf B247} (1984),
83
\item{39.} Manin, Yu.I.: invited paper presented at the XVIII Int. Coll.
on Group Theoretical Methods in Physics, Moscow, USSR, June 4-9 1990
\item{40.} Connes, A.: Publ. Mat. IHES {\bf 62} (1985), 257
\item{41.} Mack, G., Schomerous, V.: preprint HVTMP 94-B335 (1994)
\item{42.} Mack, G., Schomerous, V.: Commun. Math. Phys. {\bf 149} (1992), 513
\item{43.} Gervais, J.-L.: Commun. Math. Phys. {\bf 130} (1990), 257;
Phys. Lett. {\bf 243B} (1990), 85
\item{44.} Witten, E.: Nucl. Phys. {\bf B330} (1990), 285
\item{45.} Guadagnini, E. {\it et al.}: Phys. Lett. {\bf 235B} (1990), 275
\item{46.} Alvarez-Gaum\'e, L., Gomez, C., Sierra, G.: Nucl. Phys. {\bf 319B}
(1989), 155
\item{47.} Moore, G., Reshetikhin, N.: Nucl. Phys {\bf 328B} (1989), 557
\item{48.} Gomez, C., Sierra, G.: Phys. Lett. {\bf 240B} (1990), 149
\item{49.} Jones, V.F.R.: Bull. Amer. Math. Soc. {\bf 12} (1985), 103
\item{50.} Kauffman, L.: Int. J. Mod. Phys. {\bf 5A} (1990), 93
\item{51.} Greenberg, O.W.: in Proc. Argonne Workshop on Quantum Groups,
T. Curtright, D. Fairlie and C. Zachos Eds., World Scientific, Singapore, 1990
\item{52.} Kogan, I.I.: preprint PUPT-1439 and HEP-TH 9401093 (1994)
\item{53.} Sato, H.-T.: preprint OS-GE-40-93 and HEP-TH 9312174 (1993)
\item{54.} Truini, P., Varadarajan, V.S.: in "Symmetries in Science VI: From
the Rotation Group to Quantum Algebras", Bregenz, Austria, August 2-7 1992,
Plenum Press, New York, 1993, p.731
\item{55.} Ogievetsky, O.; Schmidke, W.B., Wess, J., Zumino, B.:
Max Plank and Berkeley preprint MPI-Ph/91-98, LBL-31703, UCB 92/04 (1991)
\item{56.} McCrimmon, K.: Bull. Amer. Math. Soc. {\bf 84} (1978), No.4, 612
\item{57.} Okubo, S.: in "Symmetries in Science VI: From
the Rotation Group to Quantum Algebras", Bregenz, Austria, August 2-7 1992,
Plenum Press, New York, 1993
\item{58.} Okubo, S.: J. Math. Phys. {\bf 34} (1993), 3273
\item{59.} Okubo, S.: J. Math. Phys. {\bf 34} (1993), 3292
\item{60.} Okubo, S.: University of Rochester Report UR-1312 (1993)
to appear in Proc. of the 15th Montreal-Rochester-Syracuse-Toronto Meeting
for High Energy Theories
\item{61.} Okubo, S.: University of Rochester Report UR-1319 (1993)
\item{62.} Okubo, S.: University of Rochester Report UR-1334 (1993)
\item{63.} Meyberg, K.: Math. Z. {\bf 115} (1970), 58
\item{64.} Koecher, M.: in "Lectures at the Rhine-Westphalia Academy
of Sciences" No. 307, Westdeutscher Verlag, Opladen, 1982, p. 53
\item{65.} Faulkner, J.R., Ferrar, J.C.: Commun. Algebra {\bf 8} (1980),
No. 11, 993
\item{66.} Zamolodchikov, A.B., Zamolodchikov, Al.B.: Ann. Phys. {\bf 120}
(1979), 253
\item{67.} Svinolupov, S.I.: Commun. Math. Phys. {\bf 143} (1992), 559
\bye